\documentstyle[preprint,aps,psfig]{revtex}

\def\lsim{\raise0.3ex\hbox{$\;<$\kern-0.75em\raise-1.1ex
\hbox{$\sim\;$}}}
\def\gsim{\raise0.3ex\hbox{$\;>$\kern-0.75em\raise-1.1ex
\hbox{$\sim\;$}}}
\begin{document}

\begin{flushright}
TMUP-HEL-9908\\
WU-HEP-99-4\\
June 1999
\end{flushright}
\begin{center}
\Large\bf
Solar Neutrinos and Leptonic CP Violation
\end{center}
\begin{center}
Hisakazu Minakata\footnote[2]{minakata@phys.metro-u.ac.jp}\\
{\it Department of Physics, Tokyo Metropolitan University \\
1-1 Minami-Osawa, Hachioji, Tokyo 192-0397, Japan, and \\
Research Center for Cosmic Neutrinos,
Institute for Cosmic Ray Research, \\ 
University of Tokyo, Tanashi, Tokyo 188-8502, Japan}
\vskip 0.5cm
Shinji Watanabe\footnote[3]{shinji@hep.phys.waseda.ac.jp}\\
{\it Department of Physics, Waseda University\\
3-4-1 Okubo, Shinjuku-ku, Tokyo 169-8555, Japan}\\
\end{center}

\begin{abstract}
We examine the possibility of detecting effects of leptonic 
CP violation by precise measurement of the solar neutrinos 
within the framework of standard electroweak theory minimally 
extended to include neutrino masses and mixing.
We prove a ``no-go theorem'' which states that effects of CP 
violating phase disappear in the $\nu_e$ survival probability to the 
leading order in electroweak interaction. 
The effects due to the next-to-leading order correction is 
estimated to be extremely small, effectively closing the door to 
the possibility we intended to pursue.

\end{abstract}
\newpage

Exciting discovery of neutrino oscillation in atmospheric neutrino
observation \cite {SKatm} strongly suggests that neutrinos are massive. 
Then, it is likely that the leptonic CKM \cite{CKM} matrix, which is 
now called \cite {Ramond} as the Maki-Nakagawa-Sakata (MNS) 
\cite{MNS} matrix, exists so that the nature admits
CP violation in the lepton sector. The CP violation in the
lepton sector has been the topics of interests because of various
reasons; on the one hand it may inherit a key to understanding the
lepton-quark correspondence, and on the other hand it may provide
us with an intriguing mechanism for generating baryon asymmetry in
our universe \cite {FY}.

It is then important to explore the methods for measuring leptonic
CP violating phase. Neutrino oscillation has been known to be one
of the promising ways of measuring effects of CP violation 
\cite{early}. Its
effect may be detectable if the flavor mixing angles are not small.
Recently, detailed investigations have been done to investigate how
to measure CP violation in long-baseline neutrino oscillation
experiments, with particular emphasis on how to discriminate matter
effect contamination from the genuine effect of the Kobayashi-Maskawa 
phase \cite{AKS,MN,Tani,BGW,DRGH}.

We consider in this paper an alternative possibility of measuring
CP violation by precise measurement of the solar neutrino flux, 
assuming that neutrinos interact as dictated by the standard 
electroweak theory. 
Unfortunately, we will end up with the "no-go theorem". 
That is, we will show that the effects of CP violating phase 
vanish in any observables in solar neutrino experiments 
to leading order in the electroweak interaction.
While the "theorem" is not quite a theorem because its validity 
is limited to leading order, it effectively closes the door to any 
practical attempts to measure CP violating effects in solar neutrino 
experiments because the higher order effects are so small.
Or, in other word, if a single effect of CP violating phase is 
detected in a solar neutrino experiment, then it would imply the 
existence of the neutrino interactions beyond that of the standard 
model. 

The goal of our argument in the first half of this paper is to show
that the effect of CP-violating phase vanishes to first order in
electroweak interactions in the survival probability 
$P(\nu_e\rightarrow \nu_e)$. Notice that the solar neutrino observation 
is inherently a disappearance experiment; there is no way to detect 
appearance events in charged current interactions because the energy 
is well below the thresholds, and the neutral current interaction 
cannot distinguish between $\nu_\mu$ and $\nu_\tau$.

Toward the goal let us start by recalling the well known results for 
vacuum neutrino oscillation. There exists a general theorem that 
states that the measure for CP violation vanishes in the disappearance 
probability, i.e. 
\begin{equation}
P(\nu_\alpha \rightarrow \nu_\alpha) - 
P(\bar{\nu}_\alpha \rightarrow \bar{\nu}_\alpha) = 0
\end{equation}
in vacuum. Hereafter, the Greek indices label the flavor 
eigenstates; $\alpha = e, \mu, \tau$ for three-generations. 
This comes from the CPT invariance which implies that 
$P(\nu_{\alpha}\rightarrow \nu_{\beta}) = 
P(\bar{\nu_{\beta}} \rightarrow \bar{\nu_{\alpha}})$.
The lepton flavor mixing is described by the mixing matrix
$U_{\alpha i}$ that relates the flavor eigenstate $\nu_{\alpha}$
and mass eigenstate $\nu_i$ as $\nu_{\alpha} = U_{\alpha i}\nu_i$,
where i is the index for specifying mass eigenstates.

In the framework of three flavor mixing the feature implied by the
general theorem has a simple realization; the effect of CP
violation shows up in the vacuum neutrino oscillation probabilities
in the particular way as
\begin{equation}
P(\nu_{\beta}\rightarrow\nu_{\alpha}) -
P(\bar{\nu_{\beta}}\rightarrow\bar{\nu_{\alpha}})
= 4J_{\beta\alpha}\sin\left(\frac{\Delta m^2 L}{2E}\right),
\label{deltaP}
\end{equation}
for $\alpha \neq \beta$ and 
$J_{\beta\alpha}$ stands for the leptonic Jarlskog factor 
\cite {Jarlskog},
\begin{equation}
J_{\beta\alpha} = \mbox{Im}
[U_{\alpha 1}U_{\alpha 2}^*U_{\beta 1}^*U_{\beta 2}],
\end{equation}
the unique fermion rephasing invariant measure for CP violation. 
In fact, the CP-violating piece in 
$P(\nu_{\beta} \rightarrow \nu_{\alpha})$ is identical with that of 
$P(\bar{\nu}_{\beta} \rightarrow \bar{\nu}_{\alpha})$ apart from the 
sign, and is given by a half of the right-hand-side of (\ref{deltaP}). 
By unitarity the disappearance probability can be written in terms 
of the appearance probabilities as
$1-P(\nu_e \rightarrow \nu_e) = 
P(\nu_e \rightarrow \nu_\mu) + 
P(\nu_e \rightarrow \nu_{\tau})$. 
The CP-violating pieces in the right-hand-side of this equation 
cancel owing to the cyclic property of $J_{\beta\alpha}$. 
This is all about how the CP violating effect vanishes in the survival 
probabilities $P(\nu_\alpha \rightarrow \nu_\alpha)$ in vacuum 
neutrino oscillation.

The situation changes completely when the matter effect 
\cite{Wolfenstein} is taken into account. 
Since the matter is in general neither CPT nor CP
symmetric there is no general argument which enforces that CP
violation effect disappears in neutrino oscillation. Moreover, the
CP violation which shows up in neutrino conversion probabilities
contains both the fake matter and the genuine effects due to the 
Kobayashi-Maskawa phase.
Therefore, it is difficult to make general statement on how CP 
violating effects come in. 

To attack the problem we write down the evolution equation of three
flavor neutrinos in matter which is valid to leading order in
electroweak interaction:
\begin{equation}
i\frac{d}{dx} 
\left[
\begin{array}{c}
\nu_e \\ \nu_\mu \\ \nu_\tau
\end{array}
\right] 
=
\left\{U \left[
\begin{array}{ccc}
m_1^2/2E & 0 & 0 \\
0 & m_2^2 /2E & 0 \\
0 & 0 & m_3^2/2E 
\end{array}
\right] U^{+}
+
\left[
\begin{array}{ccc}
a(x) & 0 & 0 \\
0 & 0 & 0 \\
0 & 0 & 0
\end{array}
\right]\right\}
\left[
\begin{array}{c}
\nu_e \\ \nu_\mu \\ \nu_\tau
\end{array}
\right],
\label{evolution1}
\end{equation}
where $a(x) = \sqrt{2} G_F N_e(x)$ indicates the index of
refraction with $G_F$ and $N_e(x)$ being the Fermi constant and 
the electron number density, respectively.\footnote
{We note, in passing, that the extra phases which appear for Majorana 
neutrinos do not give any effects in neutrino evolution in vacuum and 
in matter because they are multiplied from right to $U$ and therefore 
drop out in (\ref{evolution1}).}

We take a particular parametrization of the mixing matrix 
\cite{CKMparam}. 
\begin{equation}
U = e^{i\lambda_7 \theta_{23}} \Gamma_{\delta}e^{i \lambda_5\theta_{13}}
e^{i\lambda_2\theta_{12}} 
\label{CKM}
\end{equation}
where $\lambda_i$ are $SU(3)$ Gell-Mann's matrix and $\Gamma$
contains the CP violating phase
\begin{equation}
\Gamma_{\delta} =
\left[
\begin{array}{ccc}
1 & 0 & 0 \\
0 & 1 & 0 \\
0 & 0 & e^{i \delta}
\end{array}
\right]
\end{equation}

We rewrite the evolution equation (\ref{evolution1}) in terms of 
the new basis defined \cite {KP} by
\begin{eqnarray}
\tilde{\nu_{\alpha}} &=& \left[
e^{-i\lambda_5\theta_{13}} \Gamma^{-1} e^{-i\lambda_7\theta_{23}}
\right]_{\alpha\beta} \nu_{\beta}. \\ \nonumber
&\equiv& (T^t)_{\alpha\beta}\nu_{\beta} 
\label{defT}
\end{eqnarray}
It reads
\begin{equation}
i\frac{d}{dx} 
\left[
\begin{array}{c}
\tilde{\nu_e} \\ \tilde{\nu_\mu} \\ \tilde{\nu_\tau}
\end{array}
\right] 
=
\left\{\frac{1}{2E}e^{i\lambda_2 \theta_{12}} \left[
\begin{array}{ccc}
m_1^2 & 0 & 0 \\
0 & m_2^2 & 0 \\
0 & 0 & m_3^2 
\end{array}
\right] e^{-i \lambda_2\theta_{12}}
+
a(x)
\left[
\begin{array}{ccc}
c_{13}^2 & 0 & c_{13}s_{13} \\
0 & 0 & 0 \\
c_{13}s_{13} & 0 & s_{13}^2
\end{array}
\right]\right\}
\left[
\begin{array}{c}
\tilde{\nu_e} \\ \tilde{\nu_\mu} \\ \tilde{\nu_\tau}
\end{array}
\right]
\label{evolution2}
\end{equation}
The CP phase $\delta$ disappears from the equation. It is due to
the specific way that the matter effect comes in; $a(x)$ only
appears in (1.1) element in the Hamiltonian matrix and therefore
the matter matrix diag $(a, 0, 0)$ is invariant under rotation in
$2-3$ space by $e^{i\lambda_{1}\theta_{23}}$. Then the rotation by
the phase matrix $\Gamma$ does nothing.

It is clear from (\ref{evolution2}) that any transition amplitudes 
computed with $\tilde{\nu}_{\alpha}$ basis is independent of the CP 
violating phase. 
Of course, it does not immediately imply that the CP violating phase
$\delta$ disappears in the physical transition amplitude $\langle
\nu_{\beta} \mid \nu_{\alpha} \rangle$. The latter is related with 
the transition amplitude defined with $\tilde{\nu}_{\alpha}$ basis as
\begin{equation}
\langle \nu_{\beta} \mid \nu_{\alpha} \rangle
=
T_{\alpha\gamma} T^*_{\beta\delta}
\langle \tilde{\nu_{\delta}} \mid \tilde{\nu_{\gamma}} \rangle
\end{equation}
where $T$ is defined in (\ref{defT}) and its explicit form in our 
prametrization (\ref{CKM}) of the mixing matrix reads
\begin{equation}
T =
\left[
\begin{array}{ccc}
c_{13} & 0 & s_{13} \\
-s_{23}s_{13}e^{i\delta} & c_{23} & s_{23}c_{13}e^{i\delta} \\
-c_{23}s_{13}e^{i\delta} & -s_{23} & c_{23}c_{13}e^{i\delta}
\end{array}
\right]
\end{equation}
It is then evident that the $\nu_e$ survival amplitude 
$\langle \nu_e \mid \nu_e \rangle$, and hence the probability, 
does not contain the CP violating phase:
\begin{equation}
\langle \nu_e \mid \nu_e \rangle
=
c_{13}^2\langle \tilde{\nu_e} \mid \tilde{\nu_e} \rangle
+ s_{13}^2\langle \tilde{\nu_{\tau}} \mid \tilde{\nu_{\tau}} \rangle
+ c_{13}s_{13}(\langle \tilde{\nu_{e}} \mid \tilde{\nu_{\tau}} \rangle
+ \langle \tilde{\nu_{\tau}} \mid \tilde{\nu_e} \rangle)
\end{equation}
We have checked that this conclusion is not specific to 
the particular parametrization of the MNS matrix, as it should not.

Notice that the same statement does not apply to the $\nu_{\mu}$
disappearance amplitude:
\begin{eqnarray}
\langle \nu_{\mu} \mid \nu_{\mu} \rangle &=&
s_{23}^2s_{13}^2\langle \tilde{\nu_e} \mid \tilde{\nu_e} \rangle
+ c_{23}^2\langle \tilde{\nu_{\mu}} \mid \tilde{\nu_{\mu}} \rangle
+ s_{23}^2c_{13}^2(\langle \tilde{\nu_{\tau}} \mid \tilde{\nu_{\tau}} \rangle
\nonumber\\
&&
- c_{13}s_{23}s_{13}(
e^{i\delta}\langle \tilde{\nu_{\mu}} \mid \tilde{\nu_{e}} \rangle
+ e^{-i\delta}\langle \tilde{\nu_{e}} \mid \tilde{\nu_{\mu}} \rangle
)
+ c_{23}s_{13}c_{13}(
e^{i\delta}\langle \tilde{\nu_{\mu}} \mid \tilde{\nu_{\tau}} \rangle
+ e^{-i\delta}\langle \tilde{\nu_{\tau}} \mid \tilde{\nu_{\mu}} \rangle
)
\nonumber\\
&&
- s_{23}^2c_{13}s_{13}(
\langle \tilde{\nu_{e}} \mid \tilde{\nu_{\tau}} \rangle
+ \langle \tilde{\nu_{\tau}} \mid \tilde{\nu_{e}} \rangle
)
\label{survivemu}
\end{eqnarray}
One clearly sees the $\delta$ dependence in the survival amplitude 
of $\mu$ neutrinos in (\ref{survivemu}). Notices also that
it's dependence is not necessarily $cos \delta$ because
\begin{equation}
\langle \tilde{\nu_{\beta}}(x) \mid \tilde{\nu_{\alpha}(0)} \rangle
\neq
\langle \nu_{\alpha}(x) \mid \nu_{\beta}(0) \rangle
\end{equation}
owing to the fact that $a(0) \neq a(x)$ in general.

We now turn to the problem of CP violating effect due to the 
next-to-leading order correction of the standard electroweak 
interaction. It was noticed by Botella, Lim, and Marciano \cite{BLM} 
that it gives rise to a small splitting between the indices of 
refraction of muon and tau neutrinos; 
the matter effect matrix in (\ref{evolution1}), diag.(a,0,0), 
becomes diag.(a,0,b) in an appropriate phase convention of the 
neutrino wave function.
With a suitable redefinition of the Fermi constant $G_F$ 
$a(x)$ is still given by the same form, $a(x) = \sqrt{2} G_F N_e(x)$, 
and the ratio $b/a$ is computed to be \cite{BLM}
\begin{eqnarray}
\displaystyle\frac{b(x)}{a(x)} &=& 
- \frac{3\alpha}{2\pi \sin^2\theta_{W}}
\left( \frac{m_{\tau}}{m_W}\right)^2 
\left[2\ln \frac{m_{\tau}}{m_W} + \frac{5}{6} 
\right] \nonumber\\
&\simeq& 5.02 \times10^{-5}
\end{eqnarray}
for an isoscalar medium where $m_{\tau}$ and $m_W$ denote the masses 
of tau lepton and W boson, respectively.
With the next-to-leading order effect the Hamiltonian matrix in 
the evolution equation (\ref{evolution2}) in $\tilde{\nu}$ basis has the 
additional term
\begin{equation}
b(x)
\left[
\begin{array}{ccc}
c_{23}^2 s_{13}^2 & c_{23}s_{23}s_{13}e^{-i\delta} & -c_{23}^2 c_{13}s_{13} \\
c_{23}s_{23}s_{13}e^{i\delta} & s_{23}^2 & -c_{23}s_{23}c_{13}e^{i\delta} \\
-c_{23}^2c_{13}s_{13} & -c_{23}s_{23}c_{13}e^{-i\delta} & c_{23}^2c_{13}^2
\end{array}
\right].
\end{equation}
Clearly the $\tilde{\nu}$ evolution is affected by the CP violating phase. 
Its effect is however small, giving rise to the small correction term 
in the flavor conversion probability which is of the order of 
$b/a \simeq 5\times10^{-5}$. 

If the neutrinos have hierarchy in their masses in such a way that the 
$\Delta m_{12}^2$ relevant for the solar neutrino conversion is
much smaller than that of the atmospheric neutrino anomaly, i,e., 
$\Delta m_{13}^2 \simeq \Delta m_{23}^2 \equiv \Delta M^2 \gg \Delta m_{12}^2$
then CP violation due to higher order 
effects are suppressed even further. 
To give a feeling we compute the next-to-leading order correction
to the $\nu_e$ survival probability under the adiabatic approximation.
We also restrict ourselves to the leading order perturbative 
correction of the effects of third generation neutrinos. 
We give here only the result, leaving the detail to ref. \cite{MW99?}. 

\begin{eqnarray}
P(\nu_e \rightarrow \nu_e) 
&=& c_{13}^4 (c_{\omega}^2 c_{12}^2 + s_{\omega}^2 s_{12}^2 ) + s_{13}^4 
\nonumber\\
\label{adiabP}
&& + \displaystyle \frac {4E}{\Delta M^2}(a-bc_{23}^2) c_{13}^2 s_{13}^2  
\left[c_{13}^2(c_{\omega}^2 c_{12}^2 + s_{\omega}^2 s_{12}^2) - s_{13}^2
\right] \\
&&+ \displaystyle \frac {4E}{\Delta M^2} b c_{13}^2 (c_{12}^2 -s_{12}^2)
c_{\omega}s_{\omega}c_{23}s_{23}c_{13}^2 s_{13}\cos\delta \nonumber
\end{eqnarray}
where  $\Delta M^2 = \Delta m_{23}^2 \simeq \Delta m_{13}^2$ 
as defined above.
The angle $\omega$ in (\ref{adiabP}) is the value of an angle at the 
solar core; 
\begin{eqnarray}
\sin {2\omega} &=& \displaystyle\frac{B}{\sqrt{B^2 + \frac{1}{4} D^2}} 
\nonumber\\
B &=& \Bigm| \displaystyle\frac{\Delta m_{12}^2}{2E} c_{12}s_{12} + 
bc_{23}s_{23}s_{13}e^{-i\delta} \Bigm| \\
D &=& \frac{\Delta m_{12}^2}{2E}(c_{12}^2 - s_{12}^2) - ac_{13}^2 - 
b(c_{23}^2s_{13}^2 - s_{23}^2) \nonumber
\end{eqnarray}
which parametrizes matter mass eigenstate in 2$\times$2 submatrix. 
Notice that the CP violating effect comes out in a form very 
reminiscent of the Jarlskog factor, 
$J = c_{12}s_{12}c_{23}s_{23}c_{13}^2 s_{13}\sin\delta$,
apart from modification by the matter effect.

As clearly seen in (\ref{adiabP}), the CP violation effect is 
suppressed, in addition to the possible smallness of angle factors,  
by two small ratios, 
\begin{equation}
\frac{b}{\frac{\Delta M^2}{2E}} \sim 
\frac{b}{a} \frac{\Delta m_{12}^2}{\Delta M^2} 
\simeq 5 \times (10^{-8} - 10^{-7})
\end{equation}
assuming that 
$\Delta m_{12}^2 = 10^{-6} - 10^{-5} eV^2$ and 
$\Delta M^2 = 10^{-3} eV^2$.

We argue that the double suppression of the CP phase effect is not 
the artifact of the adiabatic approximation. Since $b$ carries the 
dimension of mass, it must be compensated by $\Delta m^2/E$. The 
$\Delta m^2$ has to be $\Delta M^2$ because the CP violation 
must vanish in an effective two-flavor limit 
$\Delta M^2 \rightarrow \infty$.
By the similar argument one can show that the CP violating term 
in $P(\nu_\mu \rightarrow \nu_\mu)$ obtains a mild suppression factor
$\sim aE/\Delta M^2 \simeq 0.1$ in the earth crust.

In this paper we have proven that the effect of CP violating phase 
disappears from the survival probability $P(\nu_e\rightarrow \nu_e)$ 
within the framework of the standard model of electroweak interactions 
with minimal extension of including neutrino masses and mixing. 

Several concluding remarks are in order:

\noindent
(1) Unfortunately, our result implies that it is practically impossible 
to detect the effect of CP violating phase in solar neutrino measurements 
even with highest attainable accuracies.\footnote{
Of course, even if the effect turns out to be reasonablely large, we would 
have to worry about next on how to define the CP-violating observables 
to actually design the experiments.} 

\noindent
(2) Our ``no-go theorem'' and hence the similar conclusion does apply to 
long-baseline electron (anti-) neutrino disappearance experiments. 
Muon (anti-) neutrino disappearance can be a hope but again we will face 
with the question of measurable CP odd observables.

\noindent
(3) We emphasize that our result does not mean all negative; 
it implies that super high-statistics solar neutrino experiments 
in the future will provide a clean laboratory for precise determination 
of the lepton mixing angles without any uncertainties due to the leptonic 
Kobayashi-Maskawa phase.

\noindent
(4)
It is interesting to examine how our ``theorem'' would be invalidated 
by introduction of new neutrino properties which may exist in less 
conservative extention of the standard model.

The research of HM is partly supported by 
the Grant-in-Aid for Scientific Research in Priority Areas No. 11127213,
and by 
the Grant-in-Aid for International Scientific Research No. 09045036, 
Inter-University Cooperative Research,  
Ministry of Education, Science, Sports and Culture.


\end{document}